# Phase-field modeling of chemical control of polarization stability and switching dynamics in ferroelectric thin film

Ye Cao[1] and Sergei V. Kalinin[1]

[1]Center for Nanophase Materials Sciences, Oak Ridge National Laboratory, Oak Ridge, TN 37831, USA

[Abstract]

Phase-field simulation (PFS) have revolutionized the understanding of domain structure and switching behavior in ferroelectric thin films and ceramics. Generally, PFS is based on solution of a (set) of Ginzburg-Landau equations for a defined order parameter field(s) under physical boundary conditions (BCs) of fixed potential or charge. While well-matched to the interfaces in bulk materials and devices, these BCs are generally not applicable to free ferroelectric surfaces. Here, we developed a self-consistent phase-field model with boundary conditions based on electrochemical equilibria. We chose $Pb(Zr_{0.2}Ti_{0.8})O_3$ ultrathin film consisting of (001) oriented single tetragonal domain ($P_z$) as a model system, and systematically studied the effects of oxygen partial pressure, temperature and surface ion on the ferroelectric state, and compared it with the case of complete screening. We have further explored the polarization switching induced by the oxygen partial pressure and observed pronounced size effect induced by chemical screening. Our work thus helps to understand the emergent phenomena in ferroelectric thin films brought about by the electrochemical ionic surface compensations.

To whom correspondence should be addressed:
*Email: caoy@ornl.gov*
*Email: sergei2@ornl.gov*

I. INTRODUCTION

Ferroelectrics are a class of materials possessing a spontaneous polarization that can be switched between different crystallographically equivalent states by external excitations such as electric field, mechanical stress etc. In ferroelectric thin films, the discontinuities of polarization at the surfaces create a bound charge that gives rise to a depolarization field opposite to the polarization direction. This field suppresses the ferroelectricity when the film is reduced to several unit cells dimensions.[1-4] To reduce the energy of the depolarization field and stabilize the ferroelectric film, either 180˚ domain structures with antiparallel polarization stripes,[5-8] or the surface compensating free charges to the polarization bound charges[9, 10] should be created. Even more complex phenomena emerge as the result of interaction between ferroelastic domain structures and field-induced stresses and the long range elastic fields, also controlled by mechanical boundary conditions.

The need to understand phenomena in ferroelectrics led to the emergence of phase-field simulation (PFS). In the past 15 years PFS has shown its capability in modeling a variety of ferroelectric behaviors due to its mesoscale spatial resolutions close to the dimensions of ferroelectric domain structure.[11, 12] Generally PFS is based on the solution of a set of Landau-Ginzburg-Devonshire (LGD) equations for relevant order parameters complemented by physical boundary conditions (BCs). These BCs are of ultimate importance in defining non-local electrostatic and elastic fields. In traditional PFS either fixed potential or charge BCs are set, which match well to phenomena at interfaces in bulk ferroelectric materials and devices. However these BCs fail to model the ferroelectric thin film with its surface exposed to/interacted with surface chemical environment. Therefore understanding the physics of interfacial effect to the polarization stability in ferroelectric ultrathin film is essential element of phase-field models.

It has been long understood that the electronic compensating charges from metallic electrodes could effectively screen the polarization charges and stabilize the single domain state.[9, 13-16] Recent studies revealed that even in the absence of electrodes, the single domain state can be stabilized in ultrathin film by the surface ions and defects from chemical environment.[17-19] Studies indicated that the polarization orientations could affect the polarity of absorbed surface ions [20, 21] and the equilibrium surface stoichiometry.[22] On the other hand, the surface ions could reversely flip the polarity of the polarizations in ultrathin films by changing the external chemical potential.[23] It is further implied that the switching could occur in a continuous process in which the polarization was suppressed and then flipped to the opposite polarity when the chemical potential reached the coercive bias,[24] in contrast to the consensus that domain switching occurs only through domain nucleation and growth.[5, 25]

These findings motivated the study of interactions between ferroelectric order and surface chemistry. Stephenson and Highland developed a thermodynamic model [26] (S&H model) to study the ferroelectric phase transition and equilibrium phase diagram in $PbTiO_3$ (001) film with its surface in equilibrium with chemical environment. However it applies only to monodomain in equilibrium state, while the polarization gradient in multi-domain structure and the polarization phase transition dynamics are not taken into account. Thoughts then arise naturally as how are the polarization switching driven by chemical potential comparable with those by electric bias; what is the difference in switching dynamics between monodomain and multi-domain thin film; and how do the surface ion concentration, temperature, film thickness and chemical environment affect the stability and switching behavior. These concerns can hardly be addressed from pure 0D thermodynamic approach.

In this work we developed self-consistent phase-field model combined with surface chemistry to investigate the roles of surface ionic compensations in the polarization stability and switching dynamics in ferroelectric ultrathin film, with physical boundary conditions based on electrochemical equilibria. This provides a third approach in phase-field simulation beyond the traditionally fixed bias or fixed charge boundary conditions.

The paper is organized as follows. In Part II we described the general framework of phase-field model and surface chemistry; In Part III we analyzed the effect of temperature and oxygen partial pressure on polarization equilibrium and switching dynamics in single domain $Pb(Zr_{0.2}Ti_{0.8})O_3$ thin film. These results were compared with polarization switching driven by electric field from planar electrode. In Part IV we further extended the model to incorporate 180˚ and 90 ˚ domain structures and investigated the effect of surface ions on multi-domain switching. Part V and VI discussed and concluded our results.

## II. THERORETICAL MODEL

In this section we first introduce the general framework of phase-field model for ferroelectric crystals, with traditional fixed electric potential and fixed charges electrostatic boundary conditions for perfect and non-screening situations. Based on this we develop our model with chemical compensation boundary conditions. We use the Landau-Ginzburg-Devonshire (LGD) free energy functional to describe the ferroelectric thin film, and establish the electrochemical equilibrium condition to determine the correlation between surface ion concentration, oxygen partial pressure and external chemical potential on the surface. Finally, the Poisson equations close the model by describing the interactions between electric potential, ferroelectric polarization and surface ion charges.

## A. General phase-field model

In the phase-field simulations of ferroelectric phenomena, the total free energy density of the ferroelectrics is written as a function of polarization $P_i$, elastic strain $\varepsilon_{kl}$, electric field $E_i$, and the polarization gradient $\nabla P_i$,[12]

$$f = f_{land}(P_i) + f_{grad}(\nabla P_i) + f_{elas}(P_i, \varepsilon_{kl}) + f_{elec}(P_i, E_i) \tag{1}$$

where $f_{land}$, $f_{grad}$, $f_{elas}$ and $f_{elec}$ are the Landau energy density, the gradient energy density, the elastic energy density and the electrostatic energy density respectively. The Landau free energy density is written as a polynomial expansion of $P_i$ (6$^{th}$ order),[27]

$$\begin{aligned}
f_{land}(P_i) &= \alpha_i P_i^2 + \alpha_{ij} P_i^2 P_j^2 + \alpha_{ijk} P_i^2 P_j^2 P_k^2 \\
&= \alpha_1(T)\left(P_1^2 + P_2^2 + P_3^2\right) + \alpha_{11}\left(P_1^4 + P_2^4 + P_3^4\right) + \alpha_{12}\left(P_1^2 P_2^2 + P_2^2 P_3^2 + P_3^2 P_1^2\right) \\
&+ \alpha_{111}\left(P_1^6 + P_2^6 + P_3^6\right) + \alpha_{112}\left[P_1^2\left(P_2^4 + P_3^4\right) + P_2^2\left(P_1^4 + P_3^4\right) + P_3^2\left(P_1^4 + P_2^4\right)\right] + \alpha_{123} P_1^2 P_2^2 P_3^2
\end{aligned} \tag{2}$$

in which $\alpha$'s are the Landau coefficients and only $\alpha_1$ is linearly dependent on temperature and obeys the Curie-Weiss law,

$$\alpha_1 = \frac{1}{2\varepsilon_0 C}(T - T_0) \tag{3}$$

where $\varepsilon_0$ is the vacuum permittivity, $C$ is the Curie constant and $T_0$ is the transition temperature. The gradient energy density is introduced through the polarization gradient,

$$f_{grad}(\nabla P_i) = \frac{1}{2} g_{ijkl} \left( \frac{\partial P_i}{\partial x_j} \frac{\partial P_k}{\partial x_l} \right)$$

$$= \frac{1}{2} g_{11} \left[ \left( \frac{\partial P_1}{\partial x_1} \right)^2 + \left( \frac{\partial P_2}{\partial x_2} \right)^2 + \left( \frac{\partial P_3}{\partial x_3} \right)^2 \right] + g_{12} \left( \frac{\partial P_1}{\partial x_2} \frac{\partial P_2}{\partial x_1} + \frac{\partial P_2}{\partial x_3} \frac{\partial P_3}{\partial x_2} + \frac{\partial P_3}{\partial x_1} \frac{\partial P_1}{\partial x_3} \right) \quad (4)$$

$$+ \frac{1}{2} g_{44} \left[ \left( \frac{\partial P_1}{\partial x_2} + \frac{\partial P_2}{\partial x_1} \right)^2 + \left( \frac{\partial P_2}{\partial x_3} + \frac{\partial P_3}{\partial x_2} \right)^2 + \left( \frac{\partial P_3}{\partial x_1} + \frac{\partial P_1}{\partial x_3} \right)^2 \right]$$

in which $g_{ijkl}$ are the gradient energy coefficient tensor. The elastic energy density is written as,[28]

$$f_{elas} = \frac{1}{2} c_{ijkl} (\varepsilon_{ij} - \varepsilon_{ij}^0)(\varepsilon_{kl} - \varepsilon_{kl}^0) \quad (5)$$

where $c_{ijkl}$ is the elastic stiffness tensor, $\varepsilon_{ij}$ is the total strain and $\varepsilon_{ij}^0$ is the eigenstrain induced by the spontaneous polarization $P_i$. Details of elastic energy expressions can be found in literature.[28] To consider the dipole-dipole interaction during ferroelectric domain evolution, the electrostatic energy of a domain structure is introduced through,[29]

$$f_{elec}(P_i, E_i) = -P_i \left( E_i + \frac{E_i^d}{2} \right) = -(P_1 E_1 + P_2 E_2 + P_3 E_3) - \frac{1}{2}(P_1 E_1^d + P_2 E_2^d + P_3 E_3^d) \quad (6)$$

where $E_i$ and $E_i^d$ are the applied electric field and depolarization field respectively. The electric field is related to the electric potential ($\phi$) distribution through,

$$E_i = -\nabla_i \phi \quad (i = 1 \sim 3) \quad (7)$$

and $\phi$ is obtained by solving the electrostatic equilibrium (Poisson) equation,

$$\Delta\phi = \frac{\nabla \bullet P_i}{\varepsilon_0 \varepsilon_r} \tag{8}$$

where $\Delta$ is the Laplacian operator, $\nabla$ is the gradient operator, $\varepsilon_0$ and $\varepsilon_r$ are the vacuum permittivity and dielectric constant, respectively.

To describe the temporal evolution of ferroelectric polarization and domain structure, we solve the time-dependent Landau-Ginzburg-Devonshire (LGD) equations,[12]

$$\frac{\partial P_i(\bm{x},t)}{\partial t} = -L\frac{\delta F_{total}}{\delta P_i(\bm{x},t)}, i=1,2,3, \tag{9}$$

in which $\bm{x}$ is the position, $t$ is the time, $L$ is the kinetic coefficient related to the domain movement, $\delta$ is the variational derivative operator, and $F_{total} = \int_V f dV$ is the total free energy written as the volume integral of the energy density $f$.

## B. Phase-field model with chemical boundary conditions

Classically, the electrostatic equilibrium equation (8) and time-dependent LGD equation (9) are solved with boundary conditions,

$$\phi\big|_{z=0} = 0 \text{ , and } \phi\big|_{z=L} = V_{planar} \tag{10}$$

$$\frac{\partial P_z}{\partial z}\bigg|_{z=0, L} = 0 \tag{11}$$

in which $L$ is the thickness along $z$ direction. These represent the situation when the film is subjected to top metallic electrode at fixed external bias ($V_{\text{planar}}$) (short-circuit BCs), which provide sufficient charge carriers to fully screen the surface polarization charges ($\partial P_z / \partial z = 0$). However perfect screening only exists in ideal situations, and in many cases the introduction of effective dielectric gap is required to provide realistic representation of the materials behavior. While the gap parameters or, more generally, physicals based form of boundary conditions are unavailable in PFS, they can be obtained by matching the mesoscopic theory to density functional theory studies.

Alternatively, few works used fixed charge boundary conditions to describe the non-screening situation for Eq. (8),

$$\left.\frac{\partial P_z}{\partial z}\right|_{z=0,\,L} = \rho \qquad (12)$$

However postulation of fixed charge is always an issue for realistic materials, since the polarization becomes highly unstable under large depolarization field due to lack of screening. Furthermore, divergence in the potential energy of the system suggests that the onset of chemistry-based potential screening can be possible.

Here we develop chemical boundary condition in more details based on S&H model,[26] which maintains electrochemical equilibria between surface compensating charge density, chemical potentials and chemical environment. For simplicity, we assume that the surface ionic charges are either excessive or deficient oxygen ions, i.e., the negatively-charged adsorbed oxygen ($O_{\text{ad}}^{2-}$) or the positively-charged oxygen vacancies ($V_O^{2+}$), albeit other electrochemical reactions are treated similarly without the loss of generality. The surface reaction involving ionic species and oxygen ($O_2$) in the chemical environment can be written as,[26]

$$\text{IonSite} + \frac{1}{n_i} O_2 = z_i e^- + \text{Ion}^{z_i} \tag{13}$$

in which $n_i$ is the number of surface ions per $O_2$, $e^-$ is the electronic free charge and $z_i$ is the charge number of surface ions. For negative adsorbed oxygen ions when $n_i = 2$ and $z_i = -2$, Eq. (13) is rewritten as,

$$V_{ad} + \frac{1}{2} O_2 + 2e^- = O_{ad}^{2-} \tag{14}$$

in which $V_{ad}$ is the vacant adsorption site for negative oxygen ion $O_{ad}^{2-}$ on the surface. When the surface ions are positive oxygen vacancies with $n_i = -2$ and $z_i = +2$, Eq. (13) yields,

$$O_O = \frac{1}{2} O_2 + 2e^- + V_O^{2+} \tag{15}$$

in which $O_O$ represent the occupied oxygen ion site on the surface. The surface ion concentration ($\theta_i$) is defined so that $\theta_i = 1$ when all the oxygen ion sites are occupied. When $\theta_i < 1$, the concentration of vacant ion sites is $1 - \theta_i$. For a general electrochemical reaction,

$$\Delta G = \Delta G^\circ + RT \ln K \tag{16}$$

in which $\Delta G^\circ$ is the standard energy formation. $R$ is the gas constant, $T$ is the temperature and $K$ is the reaction constant. In the chemical reaction involving charge transfer between electrons and ions, the driving force for the entire reaction $\Delta G$ is equal to the work done by transferring $Q$ amount of charge under external bias $V_{ex}$, i.e.,

$$\Delta G = W = -V_{ex}Q = -V_{ex}z_i N_A e_0 = -V_{ex}z_i F \qquad (17)$$

Here $V_{ex}$ is the external voltage difference between surface ion and bottom electrode where electrons reside, $N_A$ is the Avogadro's number, $e_0$ is the unit charge and $F$ is the Faraday constant. The equilibrium reaction constant $K$ for Eq. (13) can be written as,

$$K = \frac{\theta_i}{(1-\theta_i) P_{O_2}^{1/n_i}} \qquad (18)$$

in which $P_{O_2}$ is the oxygen partial pressure. Combining Eqs. (16) ~ (18) yield the mass-action equilibrium equation which establishes the correlation among surface ion concentration, electrochemical potential and oxygen partial pressure,

$$\frac{\theta_i}{1-\theta_i} = P_{O_2}^{1/n_i} \exp\left(\frac{-\Delta G^\circ - z_i e_0 V_{ex}}{k_B T}\right) \qquad (19)$$

To couple the chemical control of the surface with the bulk phase-field description, we modify the electrostatic equation (8) by considering the interaction among electric potential, ferroelectric polarization and surface ion compensation as,

$$\Delta \phi = \frac{\nabla \cdot P_i - \rho}{\varepsilon_0 \varepsilon_r} \qquad (20)$$

where $\rho$ is the local charge density consisting of both bulk charges and surface charges. We apply the chemical boundary condition, defined in such a way that the electric potential is zero at the bottom electrode, and equal to electrochemical voltage $V_{ex}$ on the top surface obtained from Eq. (19),

$$\phi|_{z=0} = 0 \text{ , and } \phi|_{z=L+\lambda} = V_{ex} \tag{21}$$

in which $L$ and $\lambda$ are the thickness of the thin film and the dielectric gap atop of the film.

The polarization boundary conditions of Eq. (9) is defined at the top surface and bottom electrode as,

$$P_z|_{z=L+\lambda} = 0 \text{ , and } \left.\frac{\partial P_z}{\partial z}\right|_{z=0} = 0 \tag{22}$$

Finally, by solving the coupled LGD equation (9), chemical equilibrium equation (19) and the modified electrostatic equilibrium equation (20), with boundary conditions (21) and (22), we can obtain both the equilibrium profiles and dynamic evolutions of ferroelectric polarization, surface ion and electric potential.

III. SINGLE DOMAIN ULTRATHIN FILM WITH IONIC COMPENSATION

As a model system we choose the prototypical (001) oriented Pb(Zr$_{0.2}$Ti$_{0.8}$)O$_3$ (PZT) ferroelectric thin film. A 3D coordinate system is set up with periodic boundary conditions along $x$ and $y$ directions, and non-periodic Chebyshev collocation boundary condition along $z$ direction. The time dependent LGD equation (7) and the Poisson equation (20) are solved using the combined semi-implicit Fourier spectral method [30] and Chebyshev collocation method. [31] The material constants and energy coefficients of PZT thin film are collected from literature.[32-34] The parameters used in the simulation are listed in Table I.

For Pb(Zr$_{0.2}$Ti$_{0.8}$)O$_3$ thin film at room temperature under compressive substrate strain, (001) oriented $c$ domain is favored. Therefore we start with a simple case of pure (001) single domain PZT thin film. Thus the model is simplified into a 1D problem, with simulation size chosen to be 1×1×128. The film thickness is assumed to be $L$, topped with a dielectric layer (either dead layer or physical dielectric gap) of thickness $\lambda \ll L$, where polarization vanishes. The surface ions are assumed to lie in a plane on top of the dielectric layer, at a distance $\lambda$ above the film surface, and remain zero elsewhere, which corresponds to the physical size of the ions.

The model geometry and 1D schematic profiles of polarization ($P_z$), surface ion concentration ($\theta_i$) and electric potential ($\phi$) are illustrated in Fig. 1. Here we assume $P_z > 0$ and the adsorbed surface ions are negative. The planes of polarization bound charge ($z = L$), the chemical surface charge ($z = L+\lambda$), and the grounded bottom electrode ($z = 0$) induce a triangle-like electric potential profile with boundaries defined from Eq. (21). To avoid singularity in the simulation, we define polarizations at the boundaries ($z = 0$ and $z = L+\lambda$) based on Eq. (22), and apply a gradient to smooth out the polarization at the surface junction ($z = L$). We also rewrite the surface ion concentration profile using a 1D Gaussian function to replace the $\delta$-function,

$$\theta(x,y,z) = \theta_i(x,y)\exp\left(-\frac{(z-(L+\lambda))^2}{2t^2}\right), \quad (0 \leq z \leq L+\lambda) \qquad (23)$$

in which $t$ controls the width of the Gaussian function, and is chosen to be 0.05 to ensure that the majorities of the ions are localized at $z = L+\lambda$. Thus the surface localized ion concentration $\theta_i(x, y)$ is converted into a 3D distribution of $\theta(x, y, z)$. Here we do not consider other charged defects in the ferroelectric bulk. Therefore the total charge density in Eq. (20) can be expressed as,

$$\rho(x, y, z) = \frac{z_i e_0 \theta(x, y, z)}{V_i} \tag{24}$$

in which $V_i$ is the bulk saturation density of ion sites. By putting $\rho(x, y, z)$ into Eq. (20), we can consider the coupling between local electric potential and space charges.

## A. Equilibrium solutions with positive/negative surface ions

We first explore the properties of the ferroelectric state of PZT film in equilibrium with positive and negative surface oxygen ions. To do this, we start from classical ferroelectric state ($P_z = P_s = 0.7 (\text{C/m}^2)$ in the film and $P_z = 0$ in the dielectric layer) and observe its evolution upon transition to chemical BC. Fig. 2 illustrates the evolutions of $P_z$ and $\phi$ in a 2.5nm thick monodomain PZT thin film under room temperature ($T$=298K) and oxygen partial pressure of $10^{-6}$ bar. When $P_z$ is positive, it generates a sheet of positive polarization bound charges at the surface of the film ($z = L$) where polarization discontinues. These positive bound charges will attract negative oxygen ions ($O_{ad}^{2-}$) ($z = L+\lambda$) for charge compensation, which is schematically illustrated in Fig. 2(a). The positive polarization bound charges and the negative surface ions create a negative (downward) electric field in the ferroelectric thin film, as well as a positive chemical potential ($V_{ex}$) atop the dielectric layer as seen from the $\phi$ profile (red solid line in Fig. 2(b)). Here we assume the bottom electrode is grounded. The polarization ($P_z$) remains constant in the ferroelectric film, and gradually decreases in the dielectric layer (red solid line in Fig. 2(c)). The external chemical potential ($V_{ex}$) is dependent on the surface oxygen ion concentration ($\theta_i$), temperature (*T*) and oxygen partial pressures ($P_{O_2}$) based on Eq. (20). When $V_{ex}$ reaches a critical coercive bias ($V_c \approx 0.1\text{V}$), the positive $P_z$ in the film is suppressed, and then flipped to the negative equilibrium state, as illustrated in Fig. 2(c).

Similar behaviors are also seen in the case of $P_z < 0$. Here the negative polarization create negative polarization bound charges which adsorb positive ions, such as oxygen vacancies ($V_O^{2+}$) to the surface (Fig. 2(d)). This causes a negative chemical potential ($V_{ex}$) atop the dielectric layer, and a positive electric field in the thin film opposite to the ferroelectric polarization orientation (Fig. 2(e)). When the chemical potential reaches $V_c$, the negative $P_z$ in the ferroelectric thin film become suppressed, and eventually switched into the positive state, as shown in Fig. 3(f). Our simulation results indicate that in both cases ($P_z > 0$ and $P_z < 0$), $P_z$ remains constant in ultrathin film and is homogeneously suppressed during the switching process; while $P_{x/y}$ components remain zero, indicating that domain nucleation does not occur. This is due to the in-plane compressive strains from the substrate that inhibit the lateral polarizations. Our simulation implies that polarization switching could occur under continuous mechanism in ultrathin film of 2.5nm, in agreement with previous experimental observations. [24]

**B. Effects of temperature and oxygen partial pressure on polarization switching**

Next we investigated the effect of chemical environment, such as oxygen partial pressures ($P_{O_2}$) on the polarization stability in 2.5nm thick PZT film with fixed surface ion concentration ($\theta_i$ =0.5) under room temperature ($T$=300K). As seen from Fig. 3(a), the averaged positive polarization in the film is ~0.6 C/m$^2$ at $P_{O_2}$ =10$^9$ bar (Stage I, black solid line). We then decreased the oxygen partial pressure from high values and found that the averaged polarization smoothly decreased and remained positive (I→II). At $P_{O_2}$ =10$^{-6}$~10$^{-7}$ bar $P_z$ flipped to the negative equilibrium state (III), and remained negative when $P_{O_2}$ further decreased. The sudden change of averaged $P_z$ in a narrow window of $P_{O_2}$ implied neither lateral polarization nucleation nor polarization rotations during switching. When $P_{O_2}$ increased from low values, the flipped (negative) polarization increased (III→IV) and switched back to positive state (IV→I) at $P_{O_2}$ = 10$^5$ bar. The dependence of $P_z$ on $P_{O_2}$ is reminiscent of the polarization-electric field (*P-E*) hysteresis loop of a classic single domain ferroelectric crystal. Here higher/lower oxygen partial pressures stabilize positive/negative polarization, in agreement with the $P_{O_2}$ - *T* phase diagrams reported by Stephenson *et. al.*.[26] Our simulations indicate that the ferroelectric polarity in an ultrathin film can be reversibly switched by manipulating the chemical environment over the film surface.

We then studied the temperature dependence of the $P_z - P_{O_2}$ hysteresis loop, as illustrated in Fig. 3(a). It was found that the hysteresis loops shrink drastically with increasing temperature, and eventually disappeared at 600K. The remnant polarization ($P_r$, defined at $P_{O_2}$ =10$^0$ bar, similar to that in the *P-E* loop) also decreased when *T* increases, since ferroelectricity was suppressed at high temperature. Another reason for the hysteresis loop shrinkage was the increase of surface chemical potential ($V_{ex}$) at elevated temperature with fixed oxygen partial pressure. Notably the temperature (600K) at which ferroelectricity disappears is even lower than the Curie temperature ($T_c$) in the bulk (752K),[32, 34] indicating that the ferroelectricity is inhibited in ultrathin film. [4]

To understand the thickness dependence of the chemical-driven polarization switching behavior, we further calculated $P_z - P_{O_2}$ hysteresis in 3.0nm and 5.0nm single domain PZT thin films under different temperatures. (Fig. 3(b) and (c)) The dependences of $P_z$ on $P_{O_2}$ at different film thickness behaved similarly. However the critical oxygen partial pressures for polarization switching, and the resulting hysteresis loop areas increased with film thickness at the same temperature. The hysteresis can still be seen in 5.0nm at 600K, while it disappeared in 2.5nm at the same temperature. The transition temperatures ($T_c$) at which hysteresis disappears were calculated to be 600K, 625K and 700K in 2.5nm, 3.0nm and 5.0nm thin film respectively, indicating that $T_c$ decreases as film becomes thinner.

**C. Phase diagram for controlled temperature and oxygen partial pressure**

To study the effect of surface ionic compensation on the ferroelectric stability, we further constructed the polarization-temperature-oxygen partial pressure ($P_z$-$T$-$P_{O_2}$) phase diagram by calculating the equilibrium $P_z$ values as a function of $T$ and $P_{O_2}$ in 2.5nm PZT thin film. It is seen from Fig. 4(a) that low $P_{O_2}$ seems to stabilize negative polarization (blue region) while high $P_{O_2}$ stabilizes positive polarization (red region). In between these two regions lies an intermediate region (green region) in which both $P_z$+ and $P_z$- can be stabilized (thus represented by $P_z \approx 0$). Aside this intermediate region, only one polarity of $P_z$ is allowed to exist. The $P_{O_2}$ window of the intermediate region spans from ~$10^{-7}$ to $10^7$ bar at 300K, becomes narrower when temperature increases and eventually disappears at 600K. This indicates that the ferroelectricity is weakened with increasing $T$ and finally behaves like dielectrics in the absence of any hysteresis. The boundaries separating the stable and intermediate regions are plotted as a function of $P_{O_2}$ and $T$ in Fig. 4(b), which converge at 600K when $P_z-$ gradually changes to $P_z+$ without $P_z+/P_z-$ coexisted state.

We also plotted the $P_z$-$T$-$P_{O_2}$ phase diagram and the phase boundaries in 5.0nm PZT thin film, as illustrated in Fig. 4(c) and (d). It is seen that single $P_z+$ or $P_z-$ stabilized region only exists at elevated temperature. When $T < 400$K, the intermediate $P_z+/P_z-$ mixed region is favored for the entire range of $P_{O_2}$ ($10^{-9}$ to $10^9$). It is also found that the phase boundaries follow exponential relations between $P_{O_2}$ and $T$, and converge at 700K when the intermediate region disappears, which is about 100K higher than that in 2.5nm thick ultrathin film. All these results indicate that ferroelectricity exists in a much wider range of oxygen partial pressure, and the ferroelectric-paraelectric phase transition occurs at a higher temperature in thicker film.

### D. Comparison between ionic and electronic compensations

The similarity between the $P_z - P_{O_2}$ (Fig. 3) and $P$-$E$ hysteresis loops aroused our interest to understand the difference between these two mechanisms. Here we assumed that the electronic carriers from the planar electrodes fully compensate the polarization bound charges at the film surface. Results were compared with the case in which the polarization charges are partially compensated by the surface ions from chemical environment. To model the surface electrode that fully compensates the polarization charges, we defined the electrostatic and polarization boundary conditions based on Eq. (10) and Eq. (11).

We assumed positive $P_z$ in 3nm thick ferroelectric thin film compensated by negative electronic charge carriers in the electrode, with no surface ions in the dielectric layer, which is schematically illustrated in Fig. 5(a). The polarization and electric potential distributions under different external bias ($V_{planar}$) are shown in Fig. 5(b) and (c). The equilibrium polarization at $V_{planar}$ =0.0V was ~0.6 C/m$^2$, and remained constant in the entire film. When external bias increased, $P_z$ decreased homogeneously throughout the film. On the other hand the electric potential linearly dropped from top to bottom electrode. The polarization switching occurred at $V_c$ ~ 0.30V. For comparison we introduced the surface ion compensation in the absence of electronic screening (Fig. 5(d)), and studied the polarization and electric potential distributions under different chemical potentials ($V_{ex}$). In Fig. 5(e) the polarization remained constant in the PZT film and diminished to 0 in the dielectric layer. The electric potential became nonlinear and reached its maximum at the ferroelectric film/dielectric layer junction ($z = L$), where uncompensated polarization bound charge arose (Fig. 5(f)). The electric potential distribution induced a negative electric field inside the film that suppressed $P_z$ even at $V_{ex}$=0.0V (black solid line in Fig. 5(e)), in comparison with the equilibrium $P_z$ distribution at zero external bias (black solid line in Fig. 5(b)). Thus it can be expected that the electric field induced from the chemical potential is higher than that from the electric bias of same magnitude. The local potentials at the film surface increased with $V_{ex}$, and finally flipped +$P_z$ into -$P_z$ when $V_{ex}$ = 0.15V, almost half of the coercive bias ($V_c$) in the electronic compensation case. Our simulation results indicate that the presence of uncompensated polarization charges destabilize the polarization in ferroelectric ultrathin film, resulting in a smaller critical bias for $P_z$ switching with chemical compensation compared with the electronic compensation.

The dependence of the switching biases ($V_c$) on film thickness is shown in Fig. 6. It is seen that $V_c$ is linearly dependent on the film thickness, in both complete electronic screening and partial ionic compensation scenarios. The coercive bias ($V_c$) with surface ion compensation (red solid line) is ~0.1V smaller than that with electronic screening (black solid line) for all thickness, due to the uncompensated polarization charges that induced an extra electric field inside the film. This implies that the interfacial effects of chemical compensation become dominant when the film thickness decreases down to several nanometers. The extrapolation of the red line to the *x*-axis (film thickness) suggests the critical film thickness (~1.5nm), at which the depolarization field makes the polarization state unstable even when $V_{ex}$=0.0V. Below 1.5nm the ferroelectricity disappears.

## IV. MULTI-DOMAIN FERROELECTRIC THIN FILM OF WITH IONIC COMPENSATION

The ferroelectric domain walls have been reported to play an important role in the domain switching dynamics. In this section we further investigated the effect of surface ions on ferroelectric domain structure and domain switching behaviors in 9nm thick $Pb(Zr_{0.2}Ti_{0.8})O_3$ (PZT) thin film. Two types of domain structures in tetragonal symmetry $Pb(Zr_{0.2}Ti_{0.8})O_3$ thin film, the antiparallel $(001)_c$ /$(00\text{-}1)_c$ domain stripes separated by 180° domain walls, as well as $(001)_c$/$(100)_a$ domain state separated by 90° twin walls are considered. We applied 2D phase-field model with periodic boundary condition along *x* (100) directions and non-periodic general boundary condition along $z$ (001) direction, with simulation size to be 64×1×64. The other parameters used in the simulation are the same as those in Section III.

## A. 180° domain structure with surface ionic compensations

The polarization switching processes in 180° striped domain structure under positive surface ions are shown in Fig. 7. Unlike the single domain case, the 180° switching was realized through domain wall motion (Fig. 7 (a) ~ (c)). Since the positive surface ions created a negative chemical potential ($V_{ex}$) and an upward electric field inside the film, $P_z$ of (001)$_c$ striped domain were almost fixed under $V_{ex}$ (Fig. 7(d)), while $P_z$ of (00-1)$_c$ striped domain remained almost constant until it flipped into positive $P_z$ state (Fig. 7(e)). It should be noted that the suppression of $P_z$ was hardly seen in (00-1)$_c$ domain stripe before switching, implying that the switching was indeed by lateral domain growth, in contrast to the continuous mechanism in single domain scenario. (Fig. 2 (c) and (f)) The electric potential ($\phi$) evolutions in (001)$_c$ and (00-1)$_c$ striped domains were shown in Fig. 7(f) and (g) respectively. The change of potential gradient in Fig. 7(f) was due to the polarization flip. The critical chemical potential ($V_c$) of polarization switching is calculated to be 0.40V (in absolute value), which is smaller than in single domain (~0.65V) due to the domain wall contribution.

We compared the switching processes with surface ionic compensation and with electronic screening from planar electrode, as illustrated in Fig. 8. In both cases 180° switching was facilitated by lateral domain wall motion and domain growth of (001)$_c$ stripped domains. Furthermore nucleation of (100)$_a$ horizontal domains were not found during the switching. The coercive bias is calculated to be 0.40V, same as the critical chemical potential ($V_c$). This indicates that the domain wall contribution is dominant during the switching processes.

**B. 90° domain structure with surface ionic compensations**

We further analyzed the surface ion effect on domain evolutions in PZT thin film consisting of $(00\text{-}1)_c$ and $(\text{-}100)_a$ domains separated by 90° twin walls, as shown in Fig. 9. Notably the $(\text{-}100)_a$ domain is slightly wider at the surface than at the bottom (Fig. 9(a)), due to the substrate constraint. On the top surface polarizations were oriented downward in the $(00\text{-}1)_c$ domains, and horizontal in the $(\text{-}100)_a$ domain. Thus positive surface ions were attracted to the film surface, which induce negative $V_{ex}$ as well as upward inner electric field. The domain evolutions under this positive electric field are illustrated in Fig. 9(b) ~ (e). The 90° domain wall motions were clearly seen (Fig. 9(b)), resulting in decreasing region of $(00\text{-}1)_c$ domain, similar to the 180° switching (Fig. 7(b)). Furthermore, a new $(001)_c$ domain was nucleated at one of the domain wall/bottom surface junctions, and grew along [110] orientations purely inside the $(\text{-}100)_a$ domain which eventually formed a vortex structure at the junction. (Fig. 9(b)) This nucleated $(001)_c$ domain continued to grow and formed 180° domain walls adjacent to the $(00\text{-}1)_c$ domains. It was found that local head-to-tail configuration was always maintained during the domain growth to avoid additional polarization charges and reduce the electrostatic energy. Thus a new $(100)_a$ triangle domain was formed between the $(001)_c$ and $(00\text{-}1)_c$ domain (Fig. 9(c)). This triangle domain continued to grow towards the bottom surface, and eventually forms $(100)_a/(001)_c$ 90° twin walls, while $(00\text{-}1)_c$ and $(\text{-}100)_a$ domain disappear (Fig. 9(d)). This structure was stable below the critical chemical potential ($V_c$) of ~0.30V, above which it further evolved into a single $(001)_c$ domain state (Fig. 9(e)).

V. DISCUSSION

While consensus has been that polarization switching in ferroelectric film can only be realized through domain nucleation and growth,[25] Highland *et. al.* recently reported that continuous switching mechanism with no domain formation is possible in sufficiently thin films (<5nm) and preferred at higher temperature, particularly in the case when the film is subjected to highly compressive substrate constraint which favors the vertical polarization to the film surface. Our model simulated the kinetic switching dynamics in 2.5nm thick PZT ultrathin film of purely monodomain, in which the polarization is homogeneously suppressed, and then flipped to the reverse equilibrium state under the critical external chemical potential of ~0.1V, thus providing further evidence to the possibility of continuous switching mechanism. Our model further indicates that in much thicker film when the intrinsic defects (such as ferroelectric and ferroelastic domain walls) are present, the switching is more likely to be facilitated by domain nucleation and growth, since the critical switching bias is largely lowered due to the contributions from domain wall (gradient) energy.

Although in the current model we used typical tetragonal PZT thin film as an example, our model can be easily extended to study other ferroelectric systems such as $BiFeO_3$ (BFO), in which the spontaneous polarization in rhombohedral symmetry is along <111> directions, neither perpendicular nor parallel to the (001) grown film surface. A super tetragonal-like domain state stabilized under large compressive strain (-4.5%) has recently been reported.[35, 36]

Thus the polarization switching in BFO thin film of purely monodomain could be accompanied by polarization rotations between different equilibrium states, and is highly dependent on the epitaxial strain conditions. Our established model thus offers opportunities to study the interaction between surface chemistry and polarization state in more complicated ferroelectric systems.

The fact that chemical environment can reversely switch the polarity of ferroelectric thin film, in a similar way to the applied voltages enables possibility to control domain state and create domain structure with less destruction. Our simulation further indicates that the critical chemical potentials for the switching with partial ionic compensation could even be lowered than the coercive bias with perfect electronic screening from surface electrodes. On the other hand the polarization orientations in the film could reversely control the charge state of the surface compensating ions. Therefore the interaction between polar state and surface ionic effect should not be neglected in the study of ferroelectric behavior in ambient environment, such as piezoresponse force microscopy in which oxygen ions and surface hydroxyl could affect measurement accuracy.[19, 37] Understanding the surface chemical effect on ferroelectric state opens up new pathways for realizing ferroelectric multi-functionalities as well as manipulating surface chemistry.

V. SUMMARY

In this work we developed coupled phase-field model and surface chemistry to study the polarization stability and switching dynamics in ferroelectric ultrathin film with its surface in equilibrium with a chemical environment that provide either excess or missing oxygen ions to compensate the surface polarization charges. We found that the ferroelectric polarity in a monodomain PZT thin film can be reversibly switched by the external chemical potential via controlling the oxygen partial pressure and surface ion concentrations. Continuous switching mechanism, in which polarization is first suppressed and then flipped to the reverse state without domain nucleation and growth, is possible in ultrathin film. And the switching bias under partial ionic screening is lowered than complete electronic screening at different thickness. Unlike single domain, the polarization switching in 180° and 90° domain structures is mainly facilitated through domain nucleation and growth. And the switching potential in multi-domain structure is even lowered due to the domain wall contribution. Our study highlights the understanding of surface chemistry effect on the ferroelectric dynamics and multi-domain interactions in a kinetic way.

Acknowledgements

This study was supported by the U.S. DOE, Office of Basic Energy Sciences (BES), Materials Sciences and Engineering Division (MSED) under FWP Grant No. ERKCZ07 (Y.C., S.V.K.). A portion of this research was conducted at the Center for Nanophase Materials Sciences, which is a DOE Office of Science User Facility.

TABLE I. Energy coefficients and parameters used in the simulation

| Landau coefficients | | Elastic coefficients | | Parameters in the simulation | |
|---|---|---|---|---|---|
| $\alpha_1 (10^8 C^{-2}m^2 N)$ | -1.485 | $c_{11}$ (GPa) | 173 | T (°C) | 25 |
| $\alpha_{11} (10^8 C^{-4}m^6 N)$ | -0.305 | $c_{12}$ (GPa) | 80.2 | System size | 1×1×128 |
| $\alpha_{12} (10^8 C^{-4}m^6 N)$ | 6.320 | $c_{44}$ (GPa) | 69.4 | Film (nm) | 2.5, 9.0 |
| $\alpha_{111} (10^8 C^{-6}m^{10} N)$ | 2.475 | $s_{11} (10^{-12} m^2 N^{-1})$ | 8.2 | $\theta_i$ | 0.5 |
| $\alpha_{112} (10^8 C^{-6}m^{10} N)$ | 0.968 | $s_{12} (10^{-12} m^2 N^{-1})$ | -2.6 | $V_i$ (cm$^{-3}$) | 6.4×10$^{-21}$ |
| $\alpha_{123} (10^9 C^{-6}m^{10} N)$ | -4.901 | $s_{44} (10^{-12} m^2 N^{-1})$ | 14.4 | $\Delta G°$ +,- (eV) | 1.0, 0.0 |
| $\alpha_{1111} (10^8 C^{-8}m^{14} N)$ | 0.0 | **Electrostrictive coefficients** | | $n_i$ +,- | -2.0, +2.0 |
| $\alpha_{1112} (10^8 C^{-8}m^{14} N)$ | 0.0 | $Q_{11} (10^{-2} C^{-2}m^4)$ | 8.1 | $z_i$ +,- | +2.0, -2.0 |
| $\alpha_{1122} (10^8 C^{-8}m^{14} N)$ | 0.0 | $Q_{12} (10^{-2} C^{-2}m^4)$ | -2.4 | $\kappa_{11}$ | 50 |
| $\alpha_{1123} (10^8 C^{-8}m^{14} N)$ | 0.0 | $Q_{44} (10^{-2} C^{-2}m^4)$ | 3.2 | $g_{11}/g_{110}$ | 0.3 |

[Figure Captions]

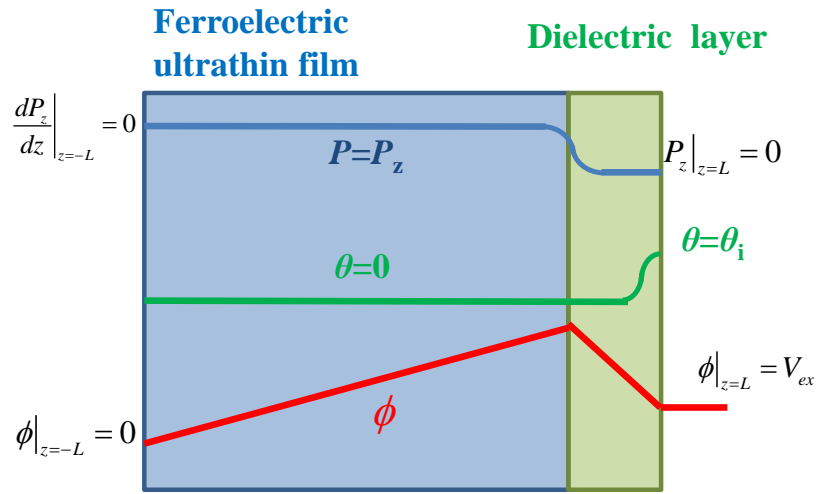

Fig. 1 Schematic plot of equilibrium profiles of ferroelectric polarization ($P_z$), ion concentration ($\theta$) and electric potential ($\phi$) in a (001) oriented ferroelectric ultrathin thin film topped with surface dielectric layer

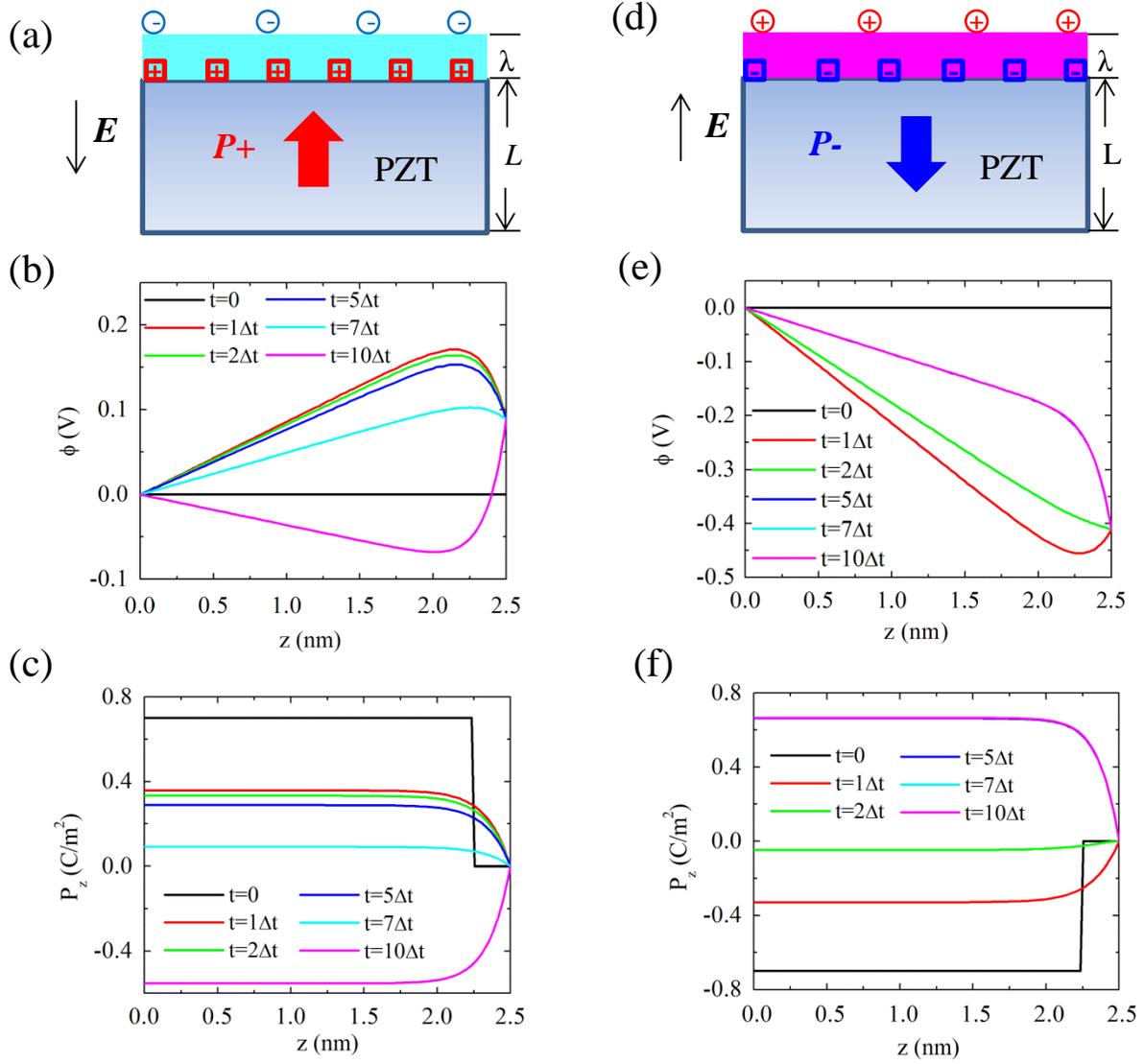

Fig. 2 Temporal evolutions of ferroelectric polarization ($P_z$) and electric potential ($\phi$) in 2.5nm thick Pb(Zr$_{0.2}$Ti$_{0.8}$)O$_3$ ultrathin film of (001) (a) ~ (c) and (00-1) (d) ~ (f) single domain at $T$=300K under constant oxygen partial pressures (10$^{-6}$ bar) and fixed surface ion concentration ($\theta_i$ =0.5). (a) and (d), Schematic illustrations of monodomain PZT thin film with polarization bound charges (squares) partially compensated by the surface ions (circles), and the induced electric field ($E$) opposite to the polarization directions; (b) and (e), electric potential evolutions; (c) and (f), polarization evolutions.

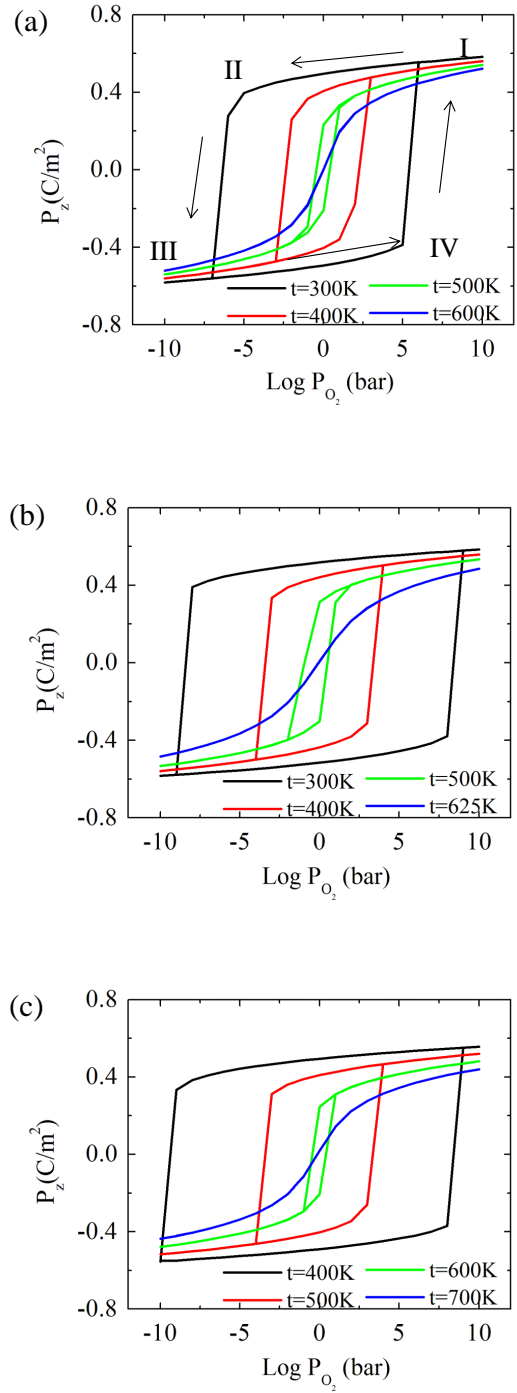

Fig. 3 Reversible switching of ferroelectric polarization ($P_z$) driven by the oxygen partial pressure ($P_{O_2}$) at different temperatures in (a) 2.5nm, (b) 3.0nm and (c) 5.0nm thick (001) PZT thin film.

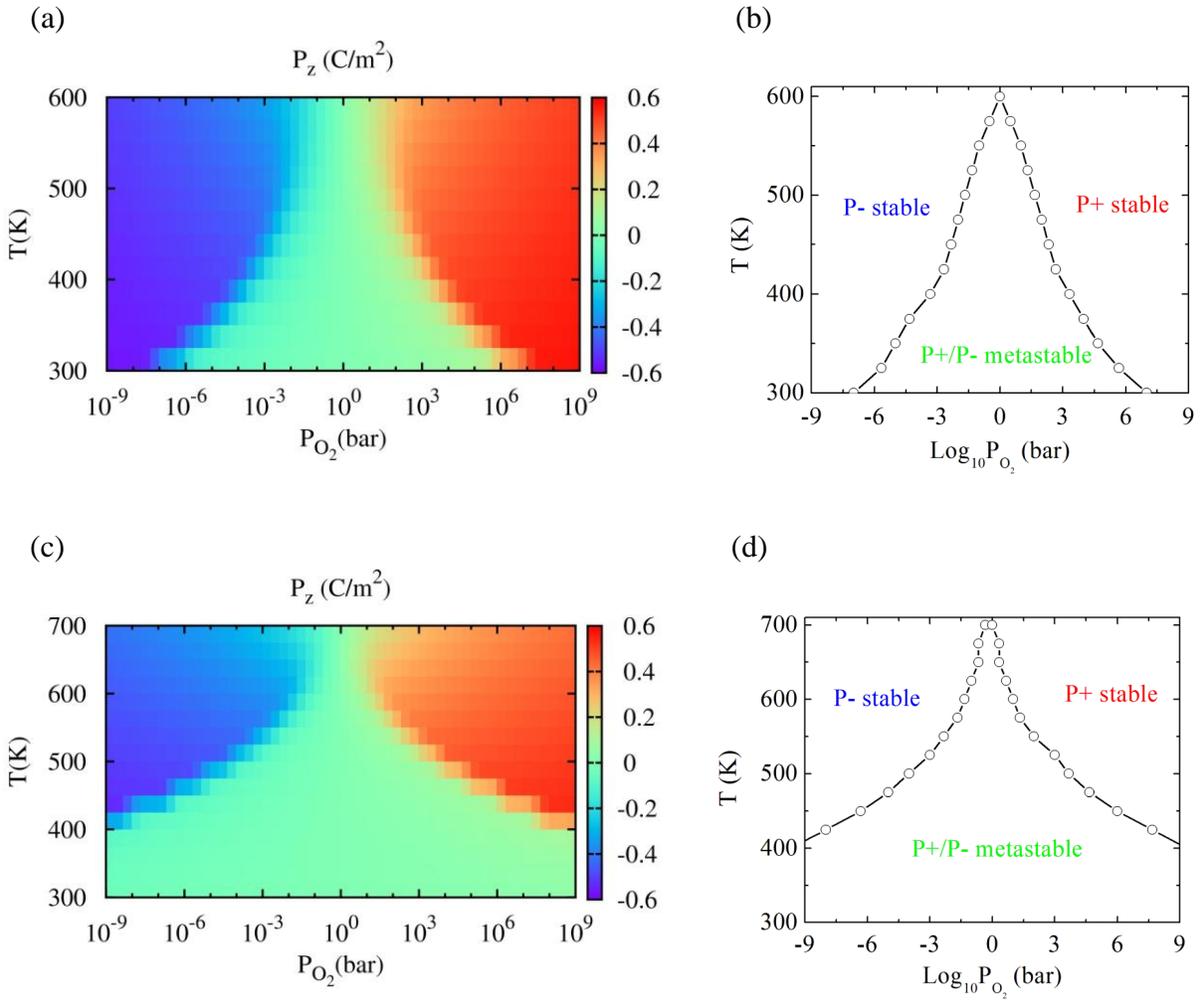

Fig. 4 Phase diagram of equilibrium polarization as a function of $P_{O_2}$ and $T$ for PZT ultrathin films of (a) 2.5nm and (c) 5.0nm; Phase boundaries separating stable and metastable polarizations in PZT ultrathin films of (b) 2.5nm and (d) 5.0nm

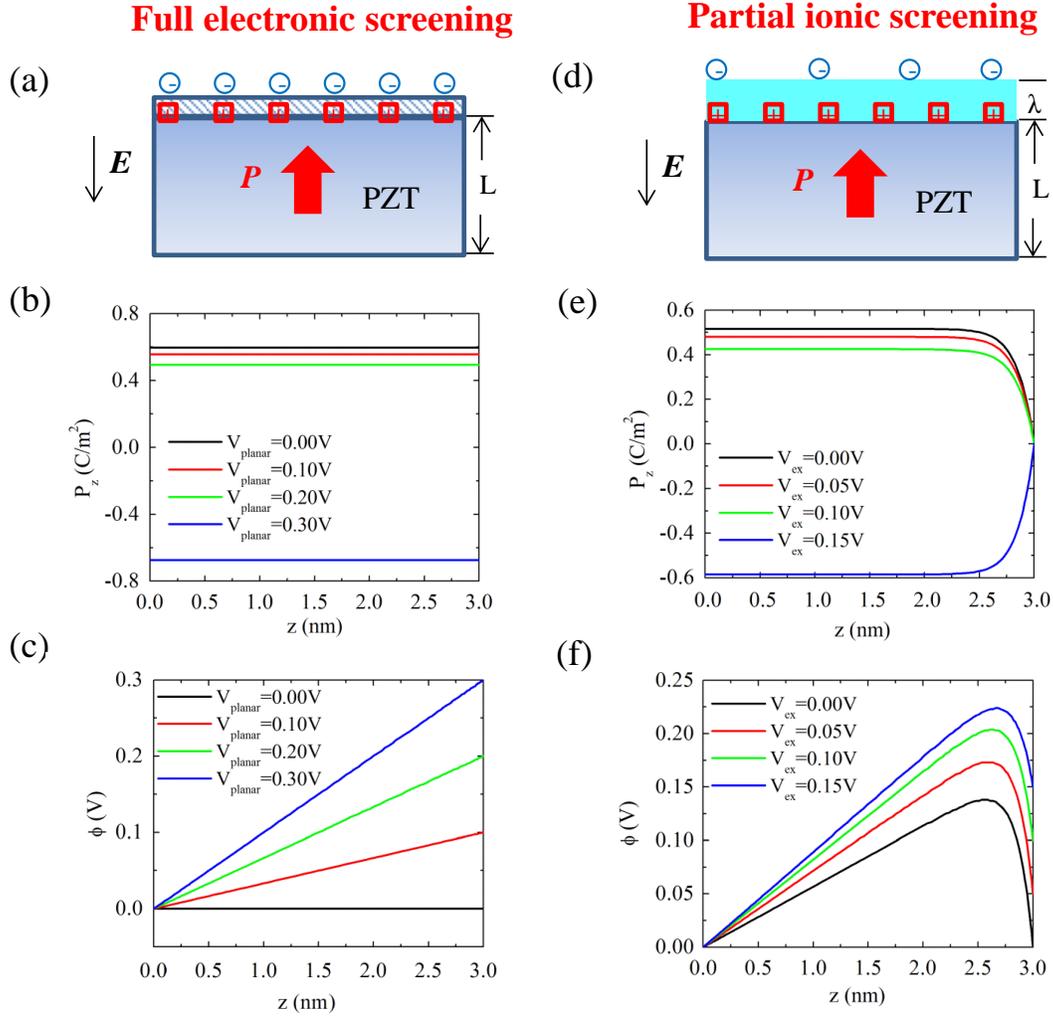

Fig. 5 Comparison of polarization ($P_z$) and electric potential ($\phi$) equilibrium profiles in 3.0 nm thick (001) PZT ultrathin film at 300K under planar electrode (a) ~ (c) and with chemical compensation (d) ~ (f) (a) Schematic plot of PZT film with atop planar electrode that provides sufficient electronic carriers to completely screen the surface polarization charges; (d) Schematic plot of PZT film exposed to chemical environment that provide negative oxygen ions to partially compensate the surface polarization charges. (b) and (c) $P_z$ and $\phi$ equilibrium profiles under different planar bias ($V_{planar}$), with the coercive bias ~0.30V; (e) and (f) $P_z$ and $\phi$ equilibrium profiles under different surface chemical potentials ($V_{ex}$), with the switching bias ~0.15V.

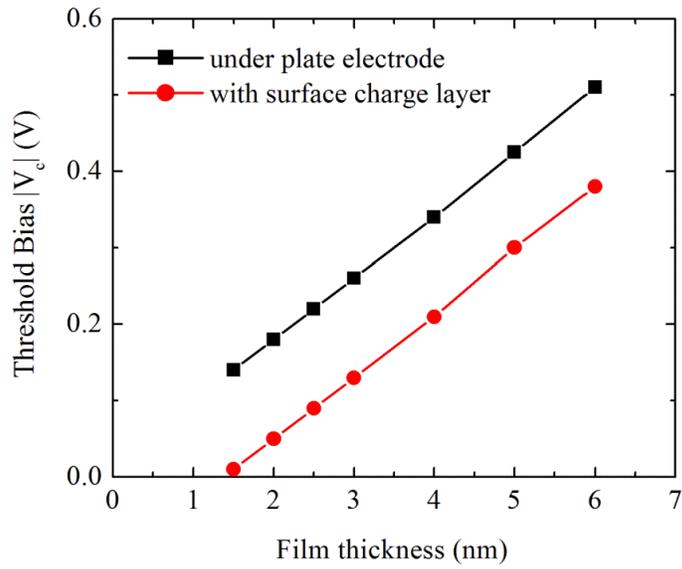

Fig. 6 Thickness dependence of switching bias in (001) PZT thin film with complete electronic screening (black solid line) and partial chemical compensation (red solid line) at 300K

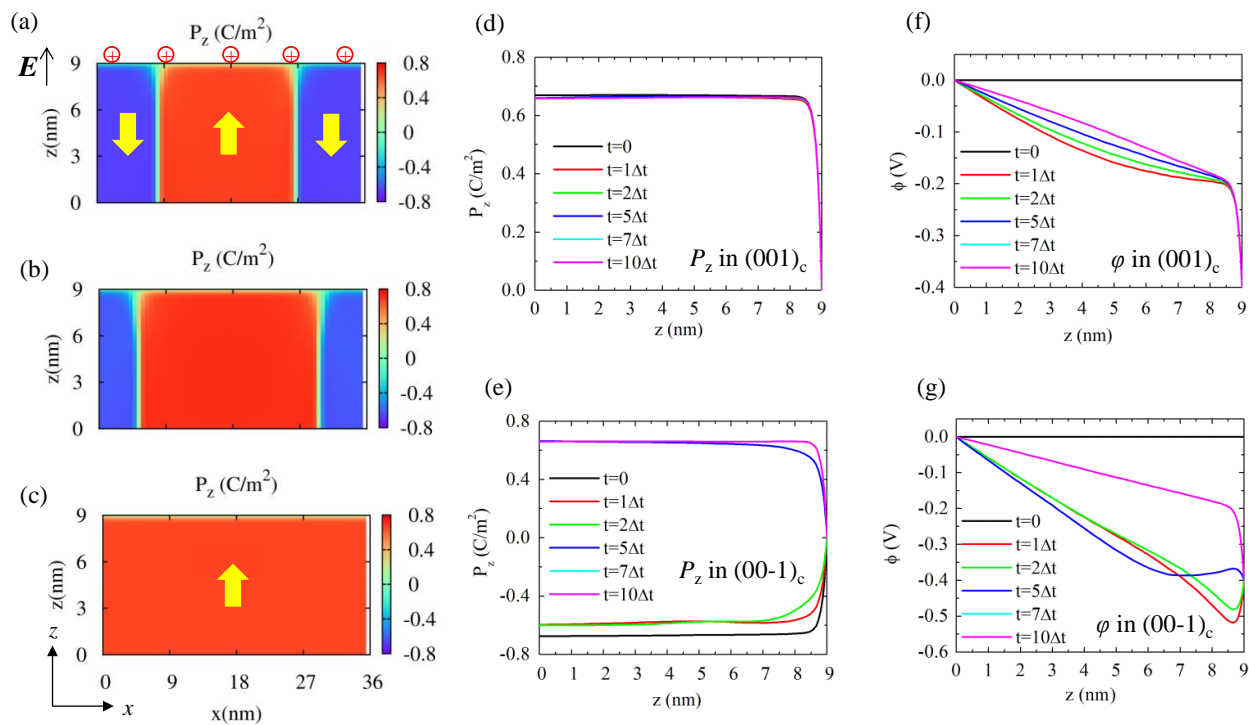

Fig. 7 Domain evolution in 9.0 nm thick PZT thin film of $(001)_c$ /$(00\text{-}1)_c$ antiparallel domain stripes with positive surface ion compensation at 300K. (a) ~ (c) 2D (*x-z*) plot domain evolution; (d) and (e), temporal evolutions of 1D polarization profiles along *z* direction in $(00\text{-}1)_c$ domain (at position *x*=4.5nm) and $(001)_c$ domain (at position *x*=18nm); (f) and (g), temporal evolutions of 1D electric potential distributions along *z* direction in $(00\text{-}1)_c$ domain (at position *x*=4.5nm) and $(001)_c$ domain (at position *x*=18nm)

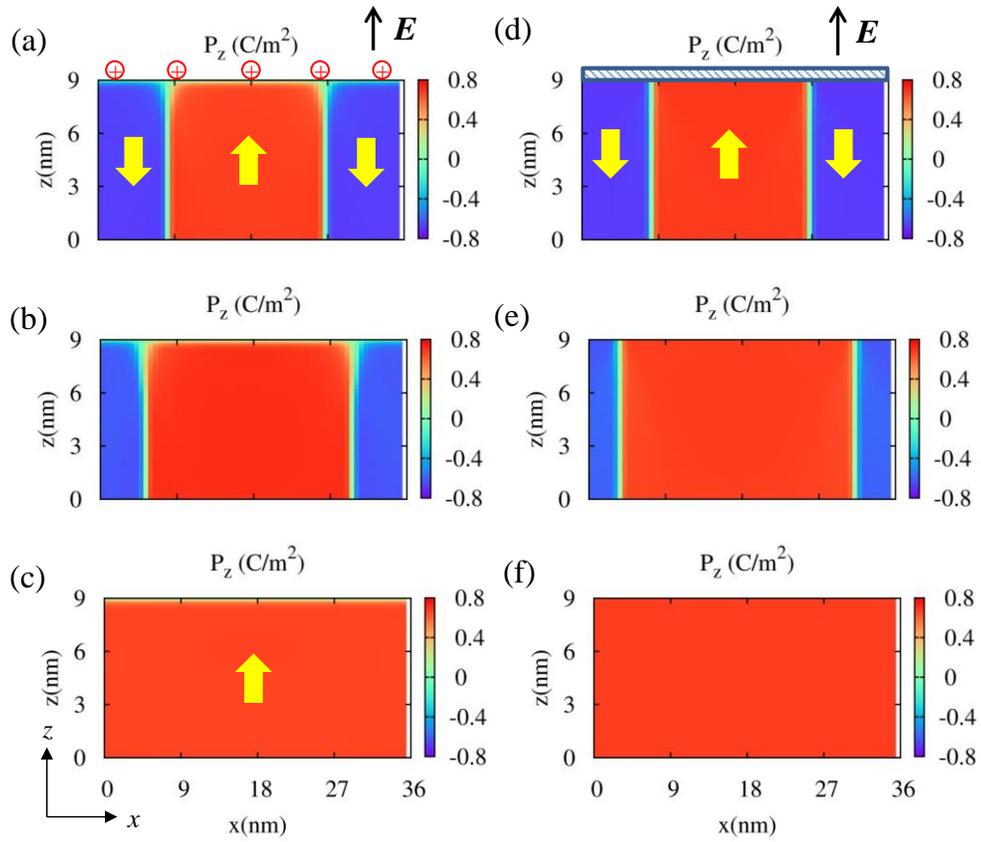

Fig. 8 Comparison of 180° domain switching in 9nm thick PZT thin film under chemical potential (a) ~ (c) and external applied electric bias (d) ~ (f) at 300K.

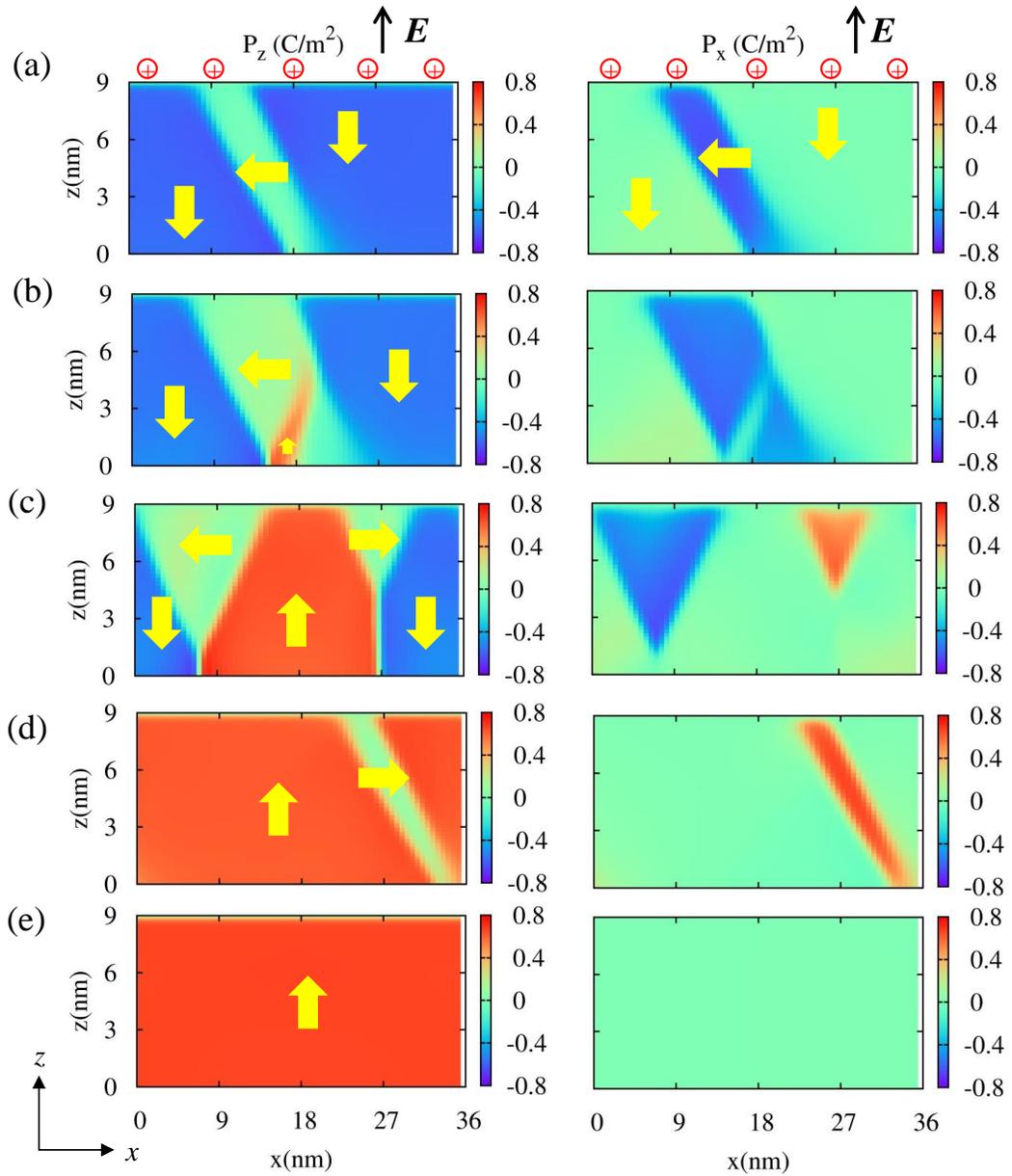

Fig. 9 Domain evolution in 9.0 nm thick PZT thin film of $(00\text{-}1)_c/(\text{-}100)_a$ 90° domain structure at 300K with positive surface ions. Left column: $P_z$ component; Right: $P_x$ component.

[References]